# Radio core dominance of *Fermi*-LAT selected AGNs

Zhi-Yuan Pei[1, 2, 3, 4], Jun-Hui Fan[3, 4], Denis Bastieri[1, 2, 3], Jiang-He Yang[3, 4] and Hu-Bing Xiao[1, 2, 3, 4]

[1] Dipartimento di Fisica e Astronomia "G. Galilei", Università di Padova, I-35131 Padova, Italy
[2] Istituto Nazionale di Fisica Nucleare, Sezione di Padova, I-35131 Padova, Italy
[3] Center for Astrophysics, Guangzhou University, Guangzhou 510006, China
[4] Astronomy Science and Technology Research Laboratory of Department of Education of Guangdong Province, Guangzhou 510006, China

2019-05-13

**ABSTRACT**

*Aims.* We present a sample of 4388 AGNs with available radio core-dominance parameters defined as the ratio of the core flux densities to extended ones, namely, $R = S_{\text{core}}/S_{\text{ext.}}$, which includes 630 *Fermi*-detected AGNs respect to the catalog of 4FGL, the fourth *Fermi* Large Area Telescope source catalog, and the rest of them are non-*Fermi*-detected AGNs. In our sample, 584 blazars are *Fermi*-detected and 1310 are not, and also consists of other subclasses such as Seyfert, Fanaroff-Riley I/II and normal galaxies. We investigate various different properties between *Fermi*-detected AGNs and non-*Fermi*-detected ones by using the core-dominance parameters as the previous study has shown that $R$ is a good indication of beaming effect.
*Methods.* We then calculate the radio spectral indices for whole sample and adopt γ-ray photon indices for *Fermi* AGNs from 4FGL catalog for discussing their different performances on different subclasses, and obtain the relation between core-dominance parameters and radio spectral indices for both *Fermi* and non-*Fermi* sources according to the two components model on radio band, which are consistent with our previous study.
*Results.* We found that the core-dominance parameters and radio spectral indices are quite different for different subclasses of AGNs, not only for *Fermi* sources but also non-*Fermi* sources, particularly, $R$ for the former ones is averagely higher than later ones. We also adopt the same relation on core-dominance parameters and γ-ray photon indices for *Fermi* sources by taking the same assumption with two components model on γ-ray band, and obtain the fitting results indicating that the γ-ray emissions of *Fermi* blazars are mainly from the core component, which is perhaps associated with the beaming effect. Therefore, *Fermi* blazars are beamed.

Use \titlerunning to supply a shorter title and/or \authorrunning to supply a shorter list of authors.

## 1. Introduction

The approximate structure of active galactic nuclei (AGNs) as we know shows that there is a supermassive black hole at the center surrounded by an accretion disk. However, the physical process of energy production and, especially, the nature of the central region, is still not clear. Based on VLA and VLBI observations of the variations in the brightness, and superluminal motion in AGNs etc., the relativistic beaming model has been constructed. VLA and VLBI observations are provided to separate the core and the extended components. Therefore, the observed radio emissions of AGNs contain two components, a compact relativistically beamed core component and an unbeamed lobe component. Blazars are the most extreme classes of AGNs are characterised by large amplitude and rapid variability, containing superluminal motion, being core dominant ($R > 1$), highly polarized, and show high energy γ-ray emission etc. All of these properties can be explained by the relativistic beaming effect. As pointing directly to the observer aside by the jet, the emission dominated by the jet is highly boosted in the line of observers sight (Urry & Padovani 1995). Multiwavelength observational data show that the radio spectrum of most blazars are usually flat: $\alpha_R = 0.00$. According to emission line features, blazars can be divided into Flat Spectrum Radio Quasars (FSRQs) and BL Lacerate objects (BL Lacs).

There several outstanding problems are proposed, such as "why some sources are γ-ray loud and others are γ-ray quite?" (Lister et al. 2015; Linford et al. 2011), "Why are some BL Lacertaes detected by Fermi, but others not?" (Wu et al. 2014b), etc. The differences between Fermi sources and non-Fermi sources have been addressed in many literature. Many authors have made great contributions here. Blazars detected by Fermi/LAT are more likely to have higher Doppler factors (e.g. Lister et al. 2009; Savolainen et al. 2010) and larger apparent opening angles (e.g. Pushkarev et al. 2009) than those not detected by Fermi/LAT. Kovalev (2009) found that the median brightness temperature values, $T_b$, for Fermi-detected sources are statistically higher than those for the rest. Savolainen et al. (2010) considered 62 objects with apparent velocity based on MOJAVE and Doppler factors, and compared the sources detected by Fermi and those not detected. They found that the Fermi-detected blazars have on average higher Doppler factors than the non-Fermi-detected blazars. Piner et al. (2012) showed that sources detected by Fermi/LAT have higher apparent speed than those not detected by Fermi/LAT. Pushkarev & Kovalev (2012) found that the Fermi AGNs have higher brightness temperature and VLBI core flux densities. Linford et al. (2012) showed that Fermi-detected BL Lacs have longer jets and are polarized than others. Kovalev (2009) have suggested that LAT-detected blazars are brighter and more luminous in the radio domain at parsec scales. Pei et al. (2016) showed that the core-dominance pa-rameter log $R$ of Fermi blazars is on average higher than non-Fermi blazars, which suggested the γ-ray sources are more radio core dominated and beamed. Xiong et al. (2015) found that there are significant differences between Fermi blazars and non-Fermi blazars for differing black hole mass, jet kinetic power, and broad-line luminosity. However, the differences in jet and accretion for Fermi blazars and non-Fermi blazars is still not clear up to now.

In a relativistic beaming model, the emission is assumed to be produced by two components, namely the beamed and the unbeamed ones. Then, the observed total emission, $S^{ob}$, is the sum of the beamed, $S^{ob}_{core}$ and unbeamed, $S_{ext.}$, emission, use $S^{ob} = S_{ext.} + S^{ob}_{core} = (1 + f\delta^p)S_{ext.}$, where $f = \frac{S^{in}_{core}}{S^{in}_{ext.}}$, $S^{in}_{core}$ is the de-beamed emission in the co-moving frame, $\delta$ is a Doppler factor, $p = \alpha + 2$ (for a continuos jet case) or $p = \alpha + 3$ (for a moving sphere case), and $\alpha$ is the spectral index ($S_\nu \propto \nu^{-\alpha}$). The ratio, $R$, of the two components is the core-dominance parameter. Some authors use the ratio of flux densities while others use the ratio of luminosities to quantify the parameter. Namely, $R = S_{core}/S_{ext.}$ or $R = L_{core}/L_{ext.}$, where $S_{core}$ or $L_{core}$ stands for core emission while $S_{ext.}$ or $L_{ext.}$ for extended emission (see Fan & Zhang (2003); Fan et al. (2011); Pei et al. (2016) and reference therein).

Doppler factor $\delta$ is the direct indicator of the jet beaming effect, therefore, a reliable determination of the Doppler factor, $\delta$, is a important way in studying the physical process associated of AGNs with the compact emission regions, etc. Actually, the radio Doppler factors are not easy to estimate, although many methods have been proposed. As this reason, we can take the core-dominance parameter of association with Doppler factor. Fan et al. (2006) found a significant correlation between log $R$ and the Doppler factor $\delta$ derived from the lowest γ-ray flux. For our discussion of beaming effect, we will adopt log $R$ instead of $\delta$.

Dermer (1995) proposed that the dependence of the γ-ray flux on the Doppler factor, which aimed to studied the different emission mechanism due to the different dependence of γ-ray flux density ($S_\gamma$). Namely, $S_\gamma \propto \delta^{2+\alpha}$ for the synchrotron self-Compton (SSC) model and $S_\gamma \propto \delta^{3+2\alpha}$ for external Compton (EC) model. If the γ-ray luminosity is taken into account, then we can expect that: $L_\gamma \propto \delta^{4+\alpha}$ for the SSC model and $L_\gamma \propto \delta^{5+2\alpha}$ for the EC model. Thus, one can use this dependence to discuss statistically the γ-ray emission mechanism. However, unfortunately, the γ-ray Doppler factors are not available for any γ-ray sample (Fan et al. 2013).

After the launching of *Fermi* Large Area Telescope, many sources are detected to be the high energy γ-ray sources. It provides us a good opportunity to study the γ-ray mechanism. The *Fermi*/LAT detects γ rays in the energy range from 20 MeV to more than 1 Tev. In the past catalogs by *Fermi*/LAT (Abdo et al. 2009, 2010b; Nolan et al. 2012; Acero et al. 2015), most of sources in these catalogs are blazars. However, we have to point out that there are far more blazars that are not detected by *Fermi*.

Based on the first eight years of science data from the *Fermi* Gamma-ray Space Telescope mission, the latest catalog, 4FGL, the fourth *Fermi* Large Area Telescope catalog of high-energy γ-ray sources, has been released, which includes 5098 sources above in the significance of 4σ, covering from 50 MeV ∓ TeV range (The Fermi-LAT collaboration 2019). Relative to the Third LAT Source Catalog (hereafter 3FGL) including 3033 sources (Acero et al. 2015), the 4FGL catalog has twice as much exposure as well as a number of analysis improvements. The data for the 4FGL catalog were taken the period into covering eight years, during the August, 2008 to August, 2008.

AGNs occupies the vast majority in 4FGL, which 3009 AGNs are included. It comprises 2938 blazars, 38 radio galaxies and 33 other AGNs. For the blazar sample, it includes 681 flat-spectrum radio quasars (FSRQs), 1102 BL Lac-type objects (BL Lacs) and 1152 blazar candidates of unknown type (BCUs)(The Fermi-LAT collaboration 2019).

Basing on Fan et al. (2011), Pei et al. (2016) compiled 1335 blazars with available core-dominance parameter, out of which 169 blazars have γ-ray emission (from 3FGL) and gave that the averaged values of core-dominance parameter, log $R$, for Fermi blazars and non-Fermi blazars are (log $R$)|$_{FB}$ = 0.99 ±0.87 and (log $R$)|$_{non-FB}$ = −0.62 ±1.15 respectively. We also calculated the averaged values of radio spectral index, $\alpha_R$ for both sample as $(\alpha_R)|_{FB} = 0.06 \pm 0.35$ and $(\alpha_R)|_{non-FB} = 0.57 \pm 0.46$. For γ-ray photon index, $\alpha^{ph}_\gamma$, we obtained that $(\alpha^{ph}_\gamma)|_{FBLL} = 2.38 \pm 0.21$ and $(\alpha^{ph}_\gamma)|_{FQ} = 2.10 \pm 0.21$ for Fermi BL Lacs and Fermi quasars, respectively. It showed that the core-dominance parameter and radio spectral index in Fermi blazars are both quite different from non-Fermi blazars with log $R$ in Fermi blazars being on average higher than non-Fermi blazars while $\alpha_R$ being on average smaller than non-Fermi blazars.

In this work, we collect 4388 AGNs with available core-dominance parameters and present the results in section 2; some discussions are given in section 3. We then conclude and summarise our findings in the final two sections. Throughout this paper, without loss of generality, we take ΛCDM model, with $\Omega_\Lambda = 0.73$, $\Omega_M = 0.27$, and $H_0 = 73 \, \text{km s}^{-1} \, \text{Mpc}^{-1}$.

## 2. SAMPLE AND RESULTS

### 2.1. Sample and Calculations

For obtaining the available core-dominance parameters, log $R$, we adopt the data in Fan et al. (2011) and Pei et al. (2019) in our sample, which included 1224 and 2400 sources, receptiviely. We also compiled a lost of relevant data from the literature of 813 sources, which are not in either Fan et al. (2011) or Pei et al. (2019) by cross-checked. For these 764 objects, we checked their identification using the NASA/IPAC EXTRAGALACTIC DATABASE (http://ned.ipac.caltech.edu/forms/byname.html) and Roma BZCAT (http://www.asdc.asi.it/bzcat/). The former gives the basic and initial identification while latter gives the classifications of BL Lacs and quasars. Furthermore, if a source is identified as "QSO" in NED, we then check whether it is identified as BL Lacs, quasars or uncertain types (herein using "unidentified") in BZCAT. If a source is identified as "G", we then only use FRI and FRII to identify the source. For the classification of Seyfert galaxy, we use "Seyfert". For those sources that are not identified as FRI, FRII or Seyfert, we use "galaxy" to label them. If the object has no identification in NED, it is also labeled as "unidentified". Therefore, we have a sample of total 4388 AGNs.

630 *Fermi*-detected AGNs are compiled in our sample, which contains 252 BL Lacs (FBLLs), 283 quasars (FQs), 49 blazars candidate uncertainties (FBCUs), and 46 *Fermi*-detected but non-blazars (FNBs). The rest of our sample, 3758 sources are non-*Fermi*-detected, which contains 198 BL Lacs (NFBLLs), 1112 quasars (NFQs), 506 Seyfert galaxies (NFSys), 1426 normal galaxies (NFGs), 340 FR type I and II galaxies (NFFRs) and 176 unidentified sources (NFU). The sample of *Fermi*-detected and non-*Fermi*-detected sources are shown in Tab. 1 and Tab. 2 receptively. The two complete tables for the whole sample of *Fermi* AGNs and non-*Fermi* AGNs are attached as an online material of this paper. In Tab.1, Col. 1 gives IAU Name; Col. 2 classification: FBLL for *Fermi* BL Lacs, FQ for *Fermi* quasars, FBCU for *Fermi* blazars candidates uncertainties and FNB for *Fermi* non-blazars; Col. 3 redshift, z; Col. 4 FL8Y name; Col 5 flux at 1 -100 GeV in unit of $cm^{-2} s^{-1}$; Col. 6 γ-ray photon index; Col. 7 core-dominance parameter compiled from Col. 11; Col. 8 luminosity distance in the unit of Mpc; Col. 9 γ-ray luminosity in the unit of erg $s^{-1}$; Col. 10 radio spectral index and Col. 11 references. In Tab.2, Col. 1 gives IAU Name; Col. 2 classification: non-FBLL for non-Fermi BL Lacs, non-FQ for non-Fermi quasars, non-FG for non-Fermi galaxies, non-FSy for non-Fermi Seyfert galaxies, non-FFRI and non-FFRII for non-Fermi FR type I and II galaxies, non-FU for non-Fermi unidentified sources; Col. 3 redshift, z; Col. 4 core-dominance parameter compiled from Col. 7; Col. 5 luminosity distance in the unit of Mpc; Col 6 radio spectral index and Col. 7 references.

In general, the observations were performed at different frequencies by various authors and studies. However, most of these data are at 5 GHz, we therefore converted the data, given in the literature at other frequencies (ν), to 5 GHz using the assumption that (Fan et al.2011;Pei et al.2016,2019)

$$S_{core}^{5\ GHz} = S_{core}^{\nu, obs},\ S_{ext.}^{5\ GHz} = S_{ext.}^{\nu, obs}(\frac{\nu}{5\ GHz})^{\alpha_{ext.}}, \quad (1)$$

then the flux densities are K-corrected, and the core-dominance parameters are finally calculated, using the expression

$$R = (\frac{S_{core}}{S_{ext.}})(1+z)^{\alpha_{core}-\alpha_{ext.}}, \quad (2)$$

In our calculation, $\alpha_{ext.}$ (or $\alpha_{unb}$) = 0.75 and $\alpha_{core}$ (or $\alpha_j$) = 0.00 are adopted (Fan et al.2011;Pei et al.2016,2019). Some data given in the literature are luminosities. If this is not at 5 GHz, we also need to convert its value to the one expected at that frequency. Then we calculated the core-dominance parameter as $\log R = \log \frac{L_{core}}{L_{ext.}}$. For flux density data, we also calculate the luminosity using $L_\nu = 4\pi d_L^2 S_\nu$, where $d_L$ stands for a luminosity distance, defined by

$$d_L = (1+z)\frac{c}{H_0}\int_1^{1+z} \frac{1}{\sqrt{\Omega_M x^3 + 1 - \Omega_M}} dx.$$

The data are mainly at 1.4 and 5 Ghz obtained from the literature, we then calculated the spectral indices, α where $S \propto \nu^{-\alpha}$. If a source has no measured redshift, then the averaged value of the corresponding group was adopted in our calculation.

For a γ-ray source, the γ-ray luminosity can be calculated from the detected photons (Abdo et al.2010a;Fan et al.2013). We assume the γ-ray photon flux associated with γ-ray photon spectrum index by a power-low spectrum, namely,

$$\frac{dN}{dE} = N_0 E^{-\alpha_\gamma^{ph}}, \quad (3)$$

where $\alpha_\gamma^{ph}$ is γ-ray photon spectral index. $N_0$ can be expressed as

$$N_0 = N_{(E_L-E_U)}\left[\frac{1}{E_L} - \frac{1}{E_U}\right], \quad (4)$$

where $N_{(E_L-E_U)}$ is the integral photons in units of photons $cm^{-2}$ $s^{-1}$ in the energy range of $E_L-E_U$. Therefore, the flux can be obtained by $f = \int_{E_L}^{E_U} EdN$, which can be expressed as

$$f = N_{(E_L-E_U)}\left[\frac{1}{E_L} - \frac{1}{E_U}\right]\ln\frac{E_U}{E_L}, \quad (6)$$

or $\alpha_\gamma^{ph} = 2$, otherwise,

$$f = N_{(E_L-E_U)}\frac{1-\alpha_\gamma^{ph}}{2-\alpha_\gamma^{ph}}\cdot\left[\frac{E_U^{2-\alpha_\gamma^{ph}}-E_L^{2-\alpha_\gamma^{ph}}}{E_U^{1-\alpha_\gamma^{ph}}-E_L^{1-\alpha_\gamma^{ph}}}\right], \quad (7)$$

in units of Gev $cm^{-2} s^{-1}$. Finally, we can obtain the K-correction γ-ray luminosity by

$$L_\gamma = 4\pi d_L^2 (1+z)^{\alpha_\gamma^{ph}-2}f, \quad (8)$$

where $d_L$ is the luminosity distance as defined above. In our calculation, we take the adoption of $E_L = 1$ GeV and $E_U = 100$ GeV.

### 2.2. Estimated Parameters and Distributions

1. Core-dominance parameters ($\log R$):

For our whole sample of 4388 AGN, we have $(\log R)|_{Total} = -0.293 \pm 1.122$, in the range from $-4.160$ to $3.918$.

For 630 *Fermi* AGNs, $(\log R)|_{FAGNs} = 0.545 \pm 1.013$, in the range from $-3.432$ to $3.918$ with the median in $0.429$; for 3758 non-*Fermi* AGNs, $(\log R)|_{NFAGNs} = -0.434 \pm 1.077$, in the range from $-4.160$ to $3.440$ with the median in $-0.429$. It shows that *Fermi* AGNs have core-dominance parameter on average higher than non-*Fermi* sources. The K-S test results indicate that the null hypothesis (they both are from the same population) cannot be rejected at the confidence level $p = 5.132 \times 10^{-81}$ ($d_{max} = 0.414$) for *Fermi* AGNs and non-*Fermi* AGNs. The distributions of core-dominance parameter, $\log R$ and cumulative probability are shown in Fig.1.

In the sample of *Fermi* detected AGNs, for 252 *Fermi* BL Lacs, $(\log R)|_{FBLL} = 0.637 \pm 0.950$, in the range from $-3.180$ to $3.918$ with the median in $0.472$; for 283 *Fermi* quasars, $(\log R)|_{FQ} = 0.706 \pm 0.914$, in the range from $-3.166$ to $3.428$ with the median in $0.580$; for 49 *Fermi* BCUs, $(\log R)|_{FBCU} = 0.119 \pm 1.339$, in the range from $-3.432$ to $2.736$ with the median in $0.217$; for 46 *Fermi* non-blazars,$(\log R)|_{FNB} = 0.493 \pm 0.808$, in the range from $-2.150$ to $1.390$ with the median in $-0.579$. The results we obtain shows that a sequence of $\log R$ for all subclasses holds: $\log R_{FQ} > \log R_{FBLL} > \log R_{FBCU} > \log R_{FNB}$. The K-S test results indicate that the null hypothesis (they both are from the same population) cannot be rejected at the following confidence level for the different sample: $p = 0.144$ ($d_{max} = 0.097$) for *Fermi* BL Lacs and *Fermi* quasars; $p = 0.061$ ($d_{max} = 0.201$) for *Fermi* BL Lacs and *Fermi* BCUs; $p = 0.010$ ($d_{max} = 0.248$) for *Fermi* quasars and *Fermi* BCUs and $p = 3.242 \times 10^{-14}$ ($d_{max} = 0.562$) for *Fermi* blazars (FBLLs + FQs + FBCUs) and *Fermi* non-blazars. The distributions of core-dominance parameter and cumulative probability are shown in Fig.2.

for $\alpha_\gamma^{ph} = 2$, otherwise,

$$N_0 = \frac{N_{(E_L - E_U)} \cdot \Sigma}{\frac{1-\alpha_\gamma^{ph}}{E_U^{1-\alpha_\gamma^{ph}} - E_L^{1-\alpha_\gamma^{ph}}}},$$ (5)

Now we turn to the averaged values of log $R$ in non-*Fermi* AGNs. For 198 non-*Fermi* BL Lacs, $(\log R)|_{\text{NFBLL}} = 0.465 \pm 0.947$, in the range from $-1.178$ to $3.350$ with the median in $0.290$; for 1112 non-*Fermi* quasars, $(\log R)|_{\text{NFQ}} = 0.032 \pm 0.871$, with the median in $0.215$; for 440 in the range from $-3.348$ to $3$.

**Table 1.** Sample of Fermi sources

| IAU Name (1) | class (2) | z (3) | 4FGL Name (4) | Flux (1 Gev-100 Gev) (5) | $\alpha_\gamma^{ph}$ (6) | log R (7) | $d_L$ (8) | log $L_\gamma$ (9) | $\alpha_R$ (10) | Ref. (11) |
|---|---|---|---|---|---|---|---|---|---|---|
| 1722+119 | FBLL | 0.018 | 4FGL J1725.0+1152 | 3.62E-09 | 1.86 | 0.48 | 5025 | 47.00 | 0.25 | Pei19 |
| 1226+023 | FQ | 0.1583 | 4FGL J1229.0+0202 | 6.26E-09 | 2.76 | 0.66 | 734 | 45.21 | 0.11 | Fan11 |
| 0045-25 | FG | 0.740 | 4FGL J0047.5-2517 | 7.83E-10 | 2.14 | -0.66 | 5025 | 46.20 | 0.82 | Pei19 |
| 0309+411B | Fsy | 0.134 | 4FGL J0313.0+4119 | 2.96E-10 | 2.56 | 0.54 | 604 | 43.76 | -0.08 | Fan11 |
| 1142+198 | FFRI | 0.0217 | 4FGL J1144.9+1937 | 2.68E-10 | 1.89 | -0.93 | 95.4 | 42.40 | 0.95 | Fan11 |
| 0433+295 | FFRII | 0.2177 | 4FGL J0437.0+2915 | 4.93E-10 | 2.38 | -2.15 | 1037 | 44.52 | 0.92 | Fan11 |
| 0705+486 | FBCU | 0.019 | 4FGL J0708.9+4838 | 9.87E-11 | 1.91 | -0.31 | 82 | 41.83 | 0.59 | Pei19 |

**Notes.** Col. 1 gives IAU Name; Col. 2 classification: FBLL for Fermi BL Lacs, FQ for Fermi quasars, FG for Fermi galaxies, FSy for Fermi Seyfert galaxies, FFRI and FFRII for Fermi FR type I and II galaxies, FBCU for Fermi blazars candidates uncertainties; Col. 3 redshift, z; Col. 4 4FGL name; Col 5 flux at 1 -100 GeV in unit of $cm^{-2} s^{-1}$; Col. 6 $\gamma$-ray photon index; Col. 7 core-dominance parameter compiled from Col. 11; Col. 8 luminosity distance in the unit of Mpc; Col. 9 $\gamma$-ray luminosity in the unit of erg $s^{-1}$; Col. 10 radio spectral index and Col. 11 references.

**Table 2.** Sample of non-Fermi sources

| Name (1) | class (2) | z (3) | log R (4) | $d_L$ (5) | $\alpha_R$ (6) | Ref. (7) |
|---|---|---|---|---|---|---|
| 1055+0519 | non-FBLL | 0.890 | 1.50 | 464 | 0.03 | Pei18 |
| 1148-001 | non-FQ | 1.980 | 0.08 | 15128 | 0.46 | Pei18 |
| 0030-219 | non-FG | 2.168 | -0.03 | 16895 | 0.75 | Pei18 |
| 1737+622 | non-FSy | 1.4402 | 0.10 | 10177 | 0.70 | Pei18 |
| 1651+049 | non-FFRI | 0.627 | -1.93 | 4646 | 0.56 | Pei18 |
| 0758+143 | non-FFRII | 1.1944 | -1.19 | 8074 | 1.00 | Fan11 |
| 1124+455 | non-FU | 0.4166 | 1.01 | 13618 | 0.26 | Fan11 |

**Notes.** Col. 1 gives IAU Name; Col. 2 classification: non-FBLL for non-Fermi BL Lacs, non-FQ for non-Fermi quasars, non-FG for non-Fermi galaxies, non-FSy for non-Fermi Seyfert galaxies, non-FFRI and non-FFRII for non-Fermi FR type I and II galaxies, non-FU for non-Fermi unidentified sources; Col. 3 redshift, z; Col. 4 core-dominance parameter compiled from Col. 7; Col. 5 luminosity distance in the unit of Mpc; Col 6 radio spectral index and Col. 7 references.

506 non-*Fermi* Seyfert galaxies, $(\log R)|_{NFS} = -0.305 \pm 0.909$, in the range from $-3.200$ to $2.229$ with the median in $-0.274$; for 1426 non-*Fermi* normal galaxies $(\log R)|_{NFG} = -0.710 \pm 1.002$, in the range from $-3.556$ to $3.077$ with the median in $-0.699$; for 340 non-*Fermi* FR type galaxies, $\log R|_{NFFR} = -1.643 \pm 0.938$, from $-4.160$ to $1.270$ with the median in $-1.469$; for 176 non-*Fermi* unidentified sources, $(\log R)|_{NFU} = -0.186 \pm 0.970$, in the range from $-2.866$ to $3.260$ with the median in $-0.212$. A sequence holds: $(\log R)|_{NFBLL} > (\log R)|_{NFQ} > (\log R)|_{NFS} > \log R |_{NFG} > \log(R_{NFFR})$. The K-S test results shows that $p = 1.173 \times 10^{-7}$ ($d_{max} = 0.221$) for non-*Fermi* BL Lacs and non-*Fermi* quasars. The distributions of core-dominance parameter and cumulative probability are shown in Fig.3.

If we take the blazars into account, for 584 *Fermi* blazars, $(\log R)|_{FB} = 0.627 \pm 0.982$, in the range from $-3.432$ to $3.918$ with the median in $0.483$; For 1310 non-*Fermi* blazars, $(\log R)|_{NFB} = 0.097 \pm 0.896$, in the range from $-3.348$ to $3.440$ with the median in $0.234$, indicating that *Fermi* blazars have core-dominance parameter on average higher than non-*Fermi* blazars. The K-S test shows that $p = 3.428 \times 10^{-31}$ ($d_{max} = 0.295$) between these two sample. The distributions of core-dominance parameter cumulative probability are shown in Fig.4.

Through the K-S test, we found that the log $R$ distributions between FAGNs and NFAGNs are significantly different (with chance probability $p = 5.132 \times 10^{-81}$), indicating that there are many different intrinsic properties between FAGNs and NFAGNs. For the subclasses, the K-S test also indicates the distributions between FQ and NFQ are significantly different (with chance probability $p = 5.00 \times 10^{-31}$). However, considering the FBLL versus NFBLL, the result from the K-S test shows that there is no significant difference (with chance probability $p = 0.00$). We consider that perhaps some objects in NFBLL sample are also the $\gamma$-ray sources but they did not be detected by *Fermi*/LAT at this moment.

2. Radio spectral index ($\alpha_R$):

From our whole sample of 3809 AGNs with adopted available radio spectral indices, $\alpha_R$, we obtain $\alpha_R|_{Total} = 0.44 \pm 0.58$, in the range from $-2.42$ to $2.53$.

For 614 *Fermi* AGNs, $(\alpha_R)|_{FAGN} = 0.125 \pm 0.475$, from $-1.550$ to $1.545$; for 3195 non-*Fermi* AGNs, $\alpha_R|_{NFAGN} = 0.500 \pm 0.573$, in the range from $-2.418$ to $2.529$. According to these results, we have that *Fermi* AGNs have radio spectral index on average smaller than non-*Fermi* AGNs. The K-S test results shows that $p = 4.518 \times 10^{-66}$ ($d_{max} = 0.383$). The distributions of radio spectral index, $\alpha_R$ and cumulative probability are shown in Fig.5.

To the subclasses of *Fermi* sources, for 243 *Fermi* BL Lacs, $\alpha_R|_{FBLL} = 0.150 \pm 0.447$, in the range from $-1.550$ to $1.480$; for 277 *Fermi* quasars, $\alpha_R|_{FQ} = 0.016 \pm 0.433$, in the range from $-1.522$ to $1.288$; for 49 *Fermi* BCUs, $\alpha_R|_{FBCU} = 0.334 \pm 0.519$, in the range from $-1.506$ to $1.324$; for 45 *Fermi* non-blazars, $\alpha_R|_{FNB} = 0.625 \pm 0.383$, in the range from $0.560$ to $1.545$. Therefore, we can have the sequence as follows: $(\alpha_R)|_{FNB} > (\alpha_R)|_{FBCU} > \alpha_R(F_{BL})| > \alpha_R(F_Q)|$. The distributions of radio spectral index and cumulative probability are shown in Fig.6.

To the subclasses of non-*Fermi* AGNs, for 182 non-*Fermi* BL Lacs, $(\alpha_R)|_{NFBLL} = 0.260 \pm 0.481$, in the range from $-1.881$ to $1.597$; for 1018 non-*Fermi* quasars, $(\alpha_R)|_{NFQ} = 0.281 \pm 0.505$, in the range from $-1.655$ to $2.529$; for 451 non-*Fermi* Seyfert galaxies, $(\alpha_R)|_{NFS} = 0.608 \pm 0.620$, in the range from −2.418 to 2.212; for 1156 non-*Fermi* normal galaxies, $(\alpha_R)|_{NFG} = 0.618 \pm 0.515$, in the range from −2.046 to 2.127; for 294 non-*Fermi* FR type galaxies, $(\alpha_R)|_{NFFR} = 0.851 \pm 0.455$, from −1.984 to 2.279; for 94 non-*Fermi* unidentified sources, $(\alpha_R)|_{NFU} = 0.280 \pm 0.959$, in the range from −2.360 to 2.425. Thus, we obtain the sequence as follow: $(\alpha_R)|_{NFFR} > (\alpha_R)|_{NFS/NFG} > (\alpha_R)|_{NFQ} > (\alpha_R)|_{NFBLL}$ (see Fig.7).

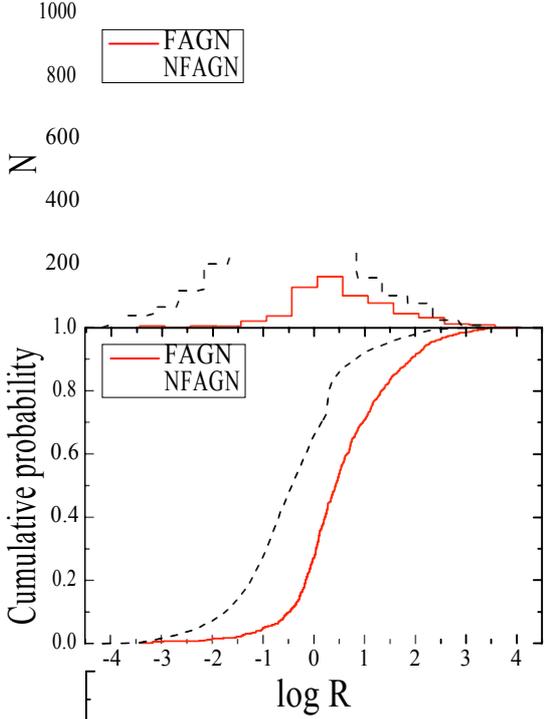
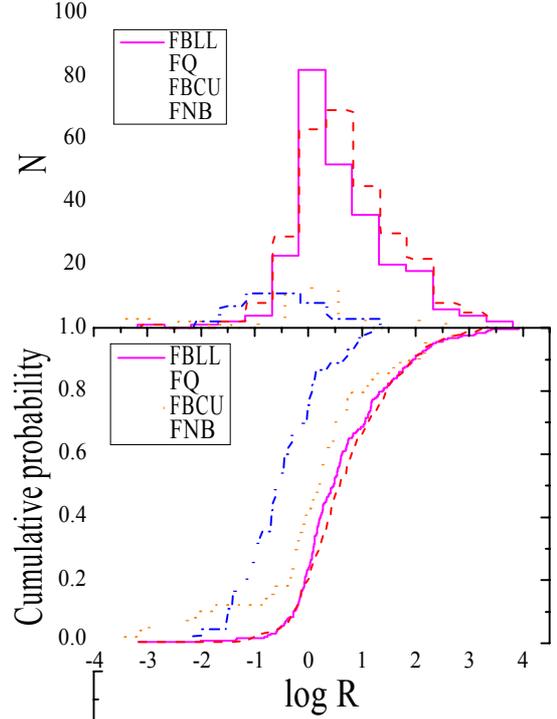

**Fig. 1.** Distribution of core-dominance parameter, log $R$ (upper panel) and cumulative probability (lower panel) for the whole sample. In this plot, red solid line stands for *Fermi* AGNs (FAGN) and black dash line for non-*Fermi* AGNs (NFAGN).

**Fig. 2.** Distribution of core-dominance parameter, log $R$ (upper panel) and the cumulative probability (lower panel) for the *Fermi*-detected AGNs. In this plot, magenta solid line stands for *Fermi* BL Lacs (FBLL), red dash line for *Fermi* quasars (FQ), orange dotted line for *Fermi* blazars candidate uncertainty sources (FBCU), blue dash-dotted line for *Fermi* non-blazars (FNB).

For 569 *Fermi* blazars, $(\alpha_R)|_{FB} = 0.085 \pm 0.459$, in the range from −1.550 to 1.480; for 1200 non-*Fermi* blazars, $(\alpha_R)|_{NFB} = 0.277 \pm 0.501$, in the range from −1.881 to 2.529, which shows that *Fermi* blazars have radio spectral index on average smaller than non-*Fermi* ones (see Fig.8).

For radio spectral index $\alpha_R$, the K-S test results indicate that the null hypothesis cannot be rejected at the following confidence level for the different samples: $p = 9.20 \times 10^{-73}$ ($d_{max} = 0.39$) for Fermi sources and non-Fermi sources; $p = 0.01$ ($d_{max} = 0.16$) for Fermi BL Lacs and non-Fermi BL Lacs; $p = 9.91 \times 10^{-22}$ ($d_{max} = 0.33$) for Fermi quasars and non-Fermi quasars; $p = 2.942 \times 10^{-17}$ ($d_{max} = 0.223$) for *Fermi* blazars and non-*Fermi* blazars. Therefore, there are significant different distributions between *Fermi* AGNs and non-*Fermi* AGNs, *Fermi* quasars and non-*Fermi* quasars but not for Fermi BL Lacs and non-Fermi BL Lacs.

3. Gamma-ray photon spectral index ($\alpha_\gamma^{ph}$):

For the whole *Fermi* AGNs, we have $(\alpha_\gamma^{ph})|_{FAGN} = 2.269 \pm 0.295$, in the range from 1.542 to 3.386 with the median in 2.2964; for 252 *Fermi* BL Lacs, we have $(\alpha_\gamma^{ph})|_{FBLL} = 2.033 \pm 0.205$, in the range from 1.578 to 2.611 with the median in 2.033; for 283 *Fermi* quasars, we have $(\alpha_\gamma^{ph})|_{FQ} = 2.473 \pm 0.181$, in the range from 1.656 to 2.969 with the median in 2.444; for 49 *Fermi* BCUs, we have $(\alpha_\gamma^{ph})|_{FBCU} = 2.276 \pm 0.283$, in the range from 1.671 to 2.772 with the median in 2.322; for 46 *Fermi* non-blazars sources, we have $(\alpha_\gamma^{ph})|_{FNB} = 2.296 \pm 0.343$, in the range from 1.542 to 3.386 with the median in 2.275; and for 584 *Fermi* blazars, we have $(\alpha_\gamma^{ph})|_{FB} = 2.267 \pm 0.291$, in the range from 1.578 to 2.969 with the median in 2.299. The distributions of gamma-ray photon spectral index, $\alpha_\gamma^{ph}$ and cumulative probability in shown in Fig.9. The K-S test results indicate that the null hypothesis cannot be rejected at the following confidence level for *Fermi* BL Lacs and *Fermi* quasars: $p = 1.974 \times 10^{-70}$ ($d_{max} = 0.775$) for *Fermi* BL Lacs and *Fermi* BCUs: $p = 1.035 \times 10^{-7}$ ($d_{max} = 0.447$), for *Fermi* quasars and *Fermi* BCUs: $p = 3.762 \times 10^{-5}$ ($d_{max} = 0.356$) and for *Fermi* blazars and *Fermi* non-blazars: $p = 0.849$ ($d_{max} = 0.094$).

4. Gamma-ray luminosity (log $L_\gamma$):

For the whole *Fermi* AGNs, we have $(\log L_\gamma)|_{FAGN} = 45.69 \pm 1.39$, in the range from 38.44 to 48.58 with the median in 45.92; for *Fermi* BL Lacs, we have $(\log L_\gamma)|_{FBLL} = 45.39 \pm 1.14$, in the range from 42.49 to 47.74 with the median in 45.42; for *Fermi* quasars, we have $(\log L_\gamma)|_{FQ} = 46.40 \pm 0.95$, in the range from 41.96 to 48.58 with the median in 46.51; for *Fermi* BCUs, we have $(\log L_\gamma)|_{FBCU} = 45.05 \pm 1.20$, in the range from 41.95 to 46.87 with the median in 45.33; Therefore, for *Fermi* blazars, we have $(\log L_{\gamma})|_{FB} = 45.85 \pm 1.19$, in the range from 41.95 to 48.58 with the median in 46.00, and for *Fermi* non-blazars, we have $(\log L_{\gamma})|_{FNB} = 43.61 \pm 1.78$, in the range from 38.44 to 46.24 with the median in 43.48.

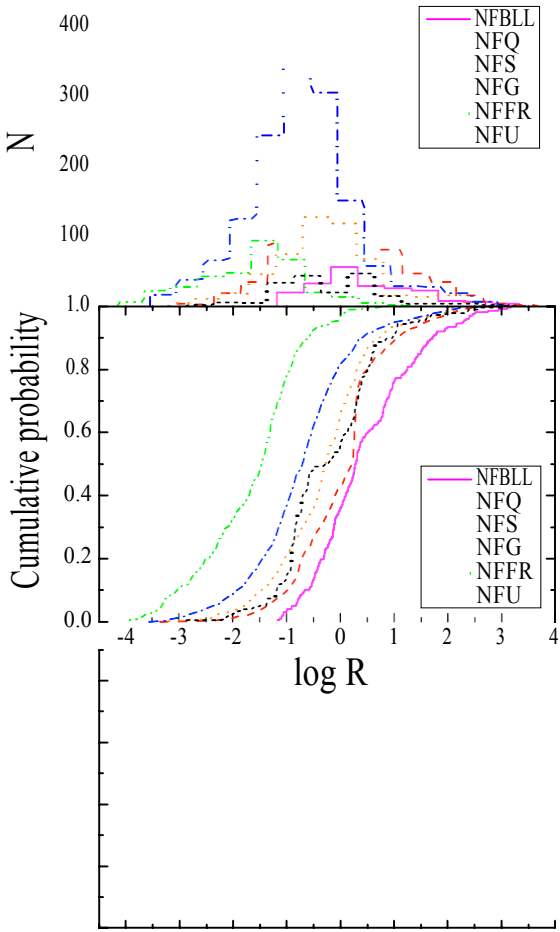
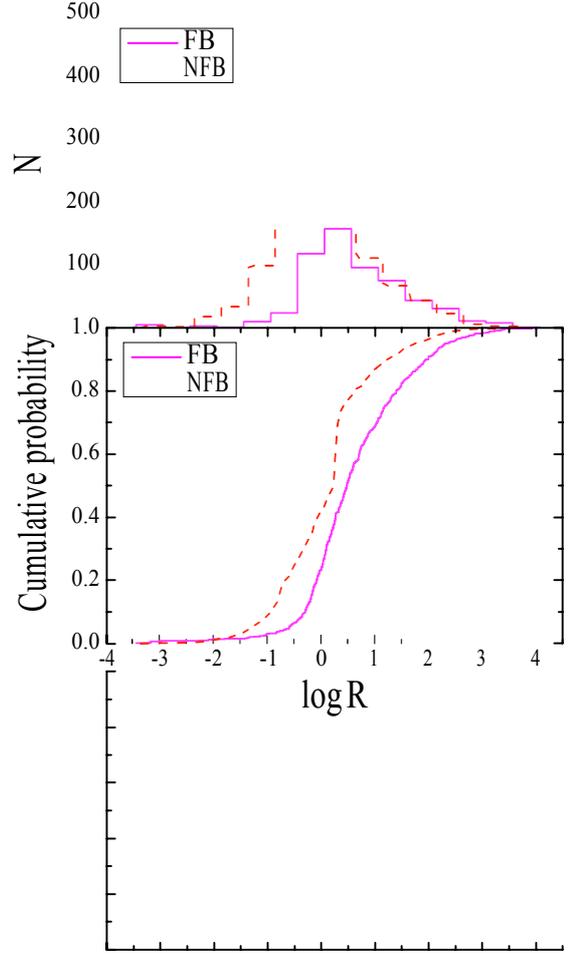

The distributions and cumulative probability in shown in Fig. 10. The K-S test shows that the probability for coming from the same population of *Fermi* BL Lacs and *Fermi* quasars is $p = 1.271 \times 10^{-20}$ ($d_{max} = 0.415$), and $p = 0.084$ ($d_{max} = 0.192$) for *Fermi* BL Lacs and *Fermi* BCUs, $p = 1.188 \times 10^{-31}$ ($d_{max} = 0.552$) for *Fermi* quasars and *Fermi* BCUs, $p = 2.460 \times 10^{-14}$ ($d_{max} = 0.565$) for *Fermi* blazars and *Fermi* non-blazars.

The statistical averaged values for our whole sample are shown in Tab.3. In this table, Col. 1 gives the sample: FS and non-FS for Fermi and non-Fermi sources, FBLL and non-FBLL for Fermi and non-Fermi BL Lacs, FQ and non-Q for Fermi and non-Fermi quasars, FG and non-G for Fermi and non-Fermi normal galaxies, FSy and non-FSy for Fermi and non-Fermi Seyfert galaxies, FFR and non-FFR for Fermi and non-Fermi FR type I and II galaxies, FBCU for Fermi blazars candidate uncertainties, non-FU for non-Fermi unidentified sources, FB and non-FB for Fermi and non-Fermi blazars; Col. 2 numbers; Col.3 averaged of core-dominance parameter; Col. 4 averaged of radio spectral index; Col. 5 averaged of γ-ray photon index for Fermi sources and Col. 6 averaged of γ-ray luminosity for Fermi sources in the

### 2.3. Parameters correlation statistics

<u>1. Correlation between redshift and core-dominance parameter:</u>

Observations of distant objects indicates that they are strong emission sources. Theoretical work reveals that such source have relativistic jet pointing to us have a large core-dominance parameter. Thus, we would expect that the core-dominance parameter should to have tendency of being positive correlated with distance (or redshift $z$).

When we study the correlation between core-dominance parameter $\log R$ and redshift $\log z$, we have $\log R = (0.21 \pm 0.08) \log z + (0.62 \pm 0.05)$ with a correlation coefficient $r = 0.11$ and a chance probability of $p = 0.00$ for *Fermi* AGNs; $\log R = (0.12 \pm 0.03) \log z - (0.38 \pm 0.02)$ with a correlation coefficient $r = 0.07$ and a chance probability of $p = 4.763 \times 10^{-5}$ for non-*Fermi* AGNs (see Fig.11).

The above correlation perhaps comes from an evolutionary explanation. For those young AGNs (low redshift $z$) have a strong beamed component, indicating the strong and uninterrupted activity, however, the extended emission has not sufficient time to accumulate, then these young sources have weak extended emission and strong beamed emission, resulting in large $\log R$ (Fan et al.2011;Pei et al.2016). Therefore, the correlation we obtain above is an approximate evolutionary effect.

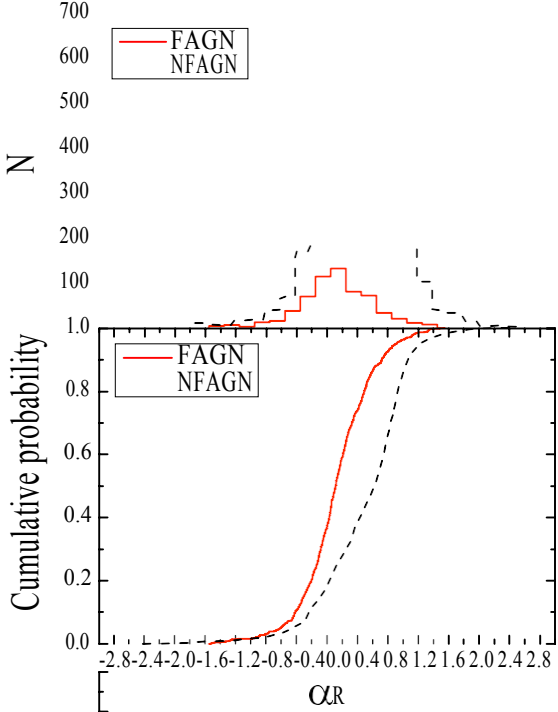
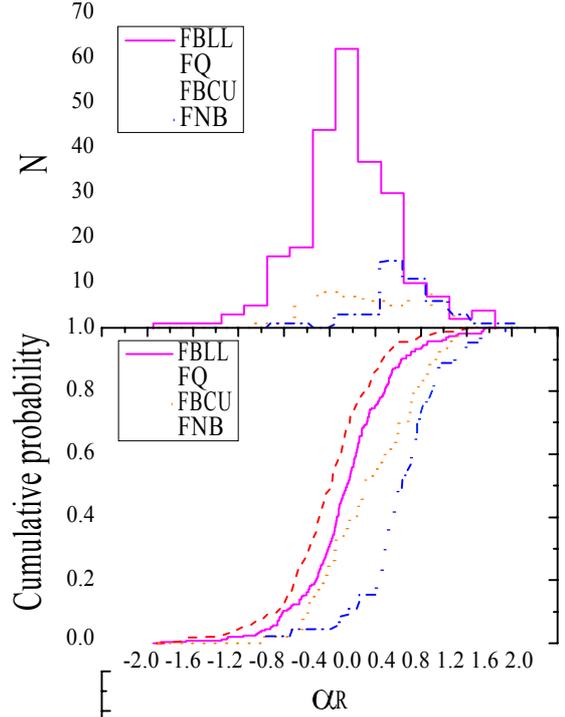

**Fig. 5.** Distribution of radio spectral index, $\alpha_R$ (upper panel) and cumulative probability (lower panel) for the whole sample. In this plot, the representations of all symbols are the same as in Fig.1.

**Fig. 6.** Distribution of radio spectral index, $\alpha_R$ (upper panel) and the cumulative probability (lower panel) for the *Fermi* AGNs. In this plot, the representations of all symbols are the same as in Fig.2.

2. Correlation between gamma-ray luminosity and core dominance parameter:

When we probe the correlation between gamma-ray luminosity ($\log L_\gamma$) and core dominance parameter ($\log R$), we can obtain that $\log L_\gamma = (0.22 \pm 0.05) \log R + (45.57 \pm 0.06)$ with with a correlation coefficient $r = 0.16$ and a chance probability of $p = 3.598 \times 10^{-5}$ for the whole *Fermi* AGNs sample (see Figure.12).

It shows that the γ-ray luminosity increases with $\log R$ for Fermi sources, indicating that that γ-ray emission is probably affected by the relativistic jet beaming effect, and $\log R$ can act as an indicator of the beaming effect.

## 3. Discussion

Blazars, and some other numbers of AGNs, shows extreme observational properties, which associated with the relativistic beaming effect. Many methods are proposed to estimate the beaming boosting factor (Ghisellini1993;Xie et al.2002,2005; Fan et al.1999). The core-dominance parameter, $R$, can be used for the orientation indicator of the jet and correlates with the polarization (Wills et al.1992;Fan et al.2006). The radio emissions in AGNs are know from the synchrotron process, which follows a correlation between radio polarization and radio spectral index. Thus, we can expect a correlation between the radio spectral index $\alpha_R$ and core-dominance parameter $\log R$.

Fan et al. (2010) gave a theoretical correlation between total radio spectral index, $\alpha_{Total}$, and core-dominance parameter $\log R$ (also see Fan et al.2011;Pei et al.2016).

$$\alpha_{Total} = \frac{R}{1+R}\alpha_{core} + \frac{1}{1+R}\alpha_{ext.}, \qquad (9)$$

We adopt this relationship to present sample and obtain the fitting results of $\alpha_{core}$ and $\alpha_{ext.}$.

As the plots in Fig.13, we can see that this relationship can not use a single simple curve to fit all of points even though it performances a clear trend for the $\alpha_R$ to be associated with $\log R$. We consider the total radio spectral index, $\alpha_{Total}$, which can also be divided into two components, namely, the core component, $\alpha_{core}$, and extended ones, $\alpha_{ext.}$, and they are different for different sources. One possibility is that the flux densities used to calculate the spectral index and those used to calculate core-dominance are not simultaneous, which results in scattering points (Fan et al.2011;Pei et al.2016). As this reason, we estimate the spectral indices, $\alpha_{core}$ and $\alpha_{ext.}$, for the whole sample by minimizing $\Sigma[\alpha_{Total} - \alpha_{core}R/(1+R) + \alpha_{ext.}(1+R)]^2$.

When we adopted this correlation across the whole sample (see Fig.13), we can get $\alpha_{core} = -0.021 \pm 0.016$ and $\alpha_{ext.} = 0.785 \pm 0.013$ with $\chi^2 = 0.257$, $R^2 = 0.220$ and a chance probability $p \theta.00$. This fitting results are consistent with the general consideration taking $\alpha_{core} = 0.00$ and $\alpha_{ext.} = 0.75$ (Fan et al.2011;Pei et al.2016).

Considering for the *Fermi* AGNs, we have $\alpha_{core} = -0.012 \pm 0.029$ and $\alpha_{ext.} = 0.402 \pm 0.048$ with $\chi^2 = 0.213$, $R^2 = 0.059$ and a chance probability $p = 1.867 \times 10^{-10}$ for the whole *Fermi*

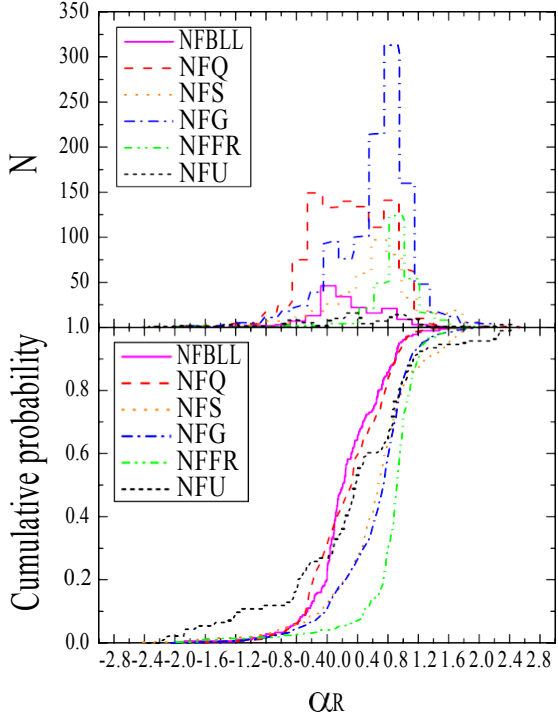

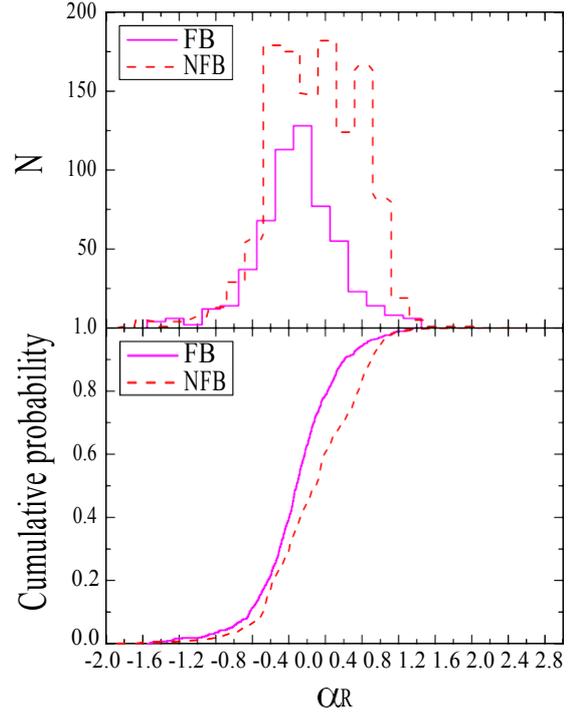

**Fig. 7.** Distribution of radio spectral index, $\alpha_R$ (upper panel) and the cumulative probability (lower panel) for the non-*Fermi* AGNs. In this plot, the representations of all symbols are the same as in Fig.3.

**Fig. 8.** Distribution of radio spectral index, $\alpha_R$ (upper panel) and cumulative probability (lower panel) for the whole blazars sample. In this plot, the representations of all symbols are the same as in Fig.4.

AGNs sample. On the other hand, for the non-*Fermi* AGNs, we have $\alpha_{core} = 0.006 \pm 0.019$ and $\alpha_{ext.} = 0.805 \pm 0.014$ with $x^2 = 0.26$, $R^2 = 0.20$ and and a chance probability $p = 0.00$ (see Fig. 14).

When we perform this relationship on all blazars in our sample, we can obtain that $\alpha_{core} = 0.104 \pm 0.035$ and $\alpha_{ext.} = 0.385 \pm 0.059$ with $x^2 = 0.210$, $R^2 = 0.024$ and a chance probability $p = 6.160 \times 10^{-11}$ for *Fermi* BL Lacs and non-*Fermi* BL Lacs (see Fig.15); $\alpha_{core} = -0.107 \pm 0.022$ and $\alpha_{ext.} = 0.643 \pm 0.027$ with $x^2 = 0.205$, $R^2 = 0.196$ and a chance probability $p \sim 0.00$ for *Fermi* quasars and non-*Fermi* quasars (also see Fig.15).

Now we turn to the investigation of correlation of $\gamma$-ray photon spectral index, $\alpha_\gamma^{ph}$, against core-dominance parameters, log $R$. As we described above, the relation (9) is associated to radio spectral indices against core-dominance parameter, and also log $R$ is well-defined at the radio band (5 GHz). Therefore, we do not have the core-dominance parameter in gamma-ray band. However, we would like to expect that there is a similar correlation between $\alpha_\gamma^{ph}$ and log $R$. For this consideration, we also use relation (9) to process our investigation. In this sence, $\alpha_{\gamma,Total}^{ph}$, $\alpha_{\gamma,core}^{ph}$ and $\alpha_{\gamma,ext.}^{ph}$ represent the $\gamma$-ray photon spectral index from total, core and extended components, respectively (see relation(10)).

$$\alpha_{\gamma, Total}^{ph} = \frac{R}{1+R}\alpha_{\gamma, core}^{ph} + \frac{1}{1+R}\alpha_{\gamma, ext.}^{ph}, \quad (10)$$

When we plot all the points of our whole sample as seen in Fig.16, we can obtain a similar curve for the sources as the correlation shown above for radio spectral index against core-dominance parameter. For the whole *Fermi* AGNs, we obtain that $\alpha_{core}^{ph} = 2.279 \pm 0.018$ and $\alpha_{ext.}^{ph} = 2.049 \pm 0.030$ with $x^2 = 0.087$, $R^2 = 0.016$ and a chance probability $p = 0.00$ (see Fig. 16). Considering the subclasses, we have $\alpha_{\gamma, core}^{ph} = 2.039 \pm 0.020$ and $\alpha_{\gamma, ext.}^{ph} = 1.821 \pm 0.038$ with $x^2 = 0.042$, $R^2 = 0.040$ and a chance probability $p = 0.00$ for the *Fermi* BL Lacs; $\alpha_{\gamma, core}^{ph} = 2.482 \pm 0.016$ and $\alpha_{\gamma, ext.}^{ph} = 2.249 \pm 0.033$ with $x^2 = 0.033$, $R^2 < 10^{-3}$ and a chance probability $p = 0.00$ for the *Fermi* quasars; $\alpha_{\gamma, core}^{ph} = 2.266 \pm 0.069$ and $\alpha_{\gamma, ext.}^{ph} = 2.488 \pm 0.084$ with $x^2 = 0.082$, $R^2 = 0.020$ and a chance probability $p = 0.00$ for the

*Fermi* blazars candidate uncertainty objects; $\alpha_{\gamma, core}^{ph} = 2.227 \pm 0.133$ and $\alpha_{\gamma, ext.}^{ph} = 2.329 \pm 0.077$ with $x^2 = 0.120$, $R^2 = 0.015$ and a chance probability $p = 0.00$ for the *Fermi* non-blazars (see Fig.17 and Tab.5).

Liu et al.(2014) indicates that the flux fluctuations of the core dominate the observed variability in the total flux density for core-dominated sources, however, the fluctuations can be suppressed by the radio jets for jet-dominated ones. Therefore, the spectrum is getting soft for high variability sources and it turns out the spectral index becomes larger. If we take the assumption that the variability comes from the central core of AGNs, than for high variability sources, e.g. Fermi blazars, the $\gamma$-ray spectral index is dominated by the core component. Our analysis shows that the core component of $\gamma$-ray photon index,

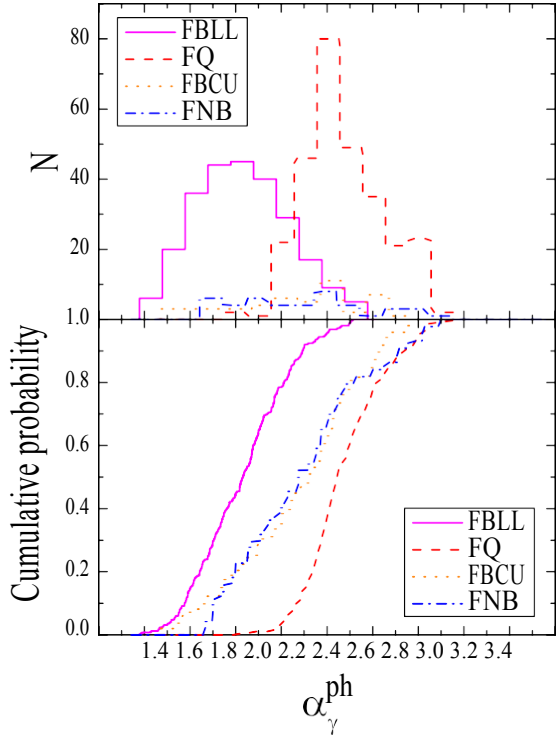 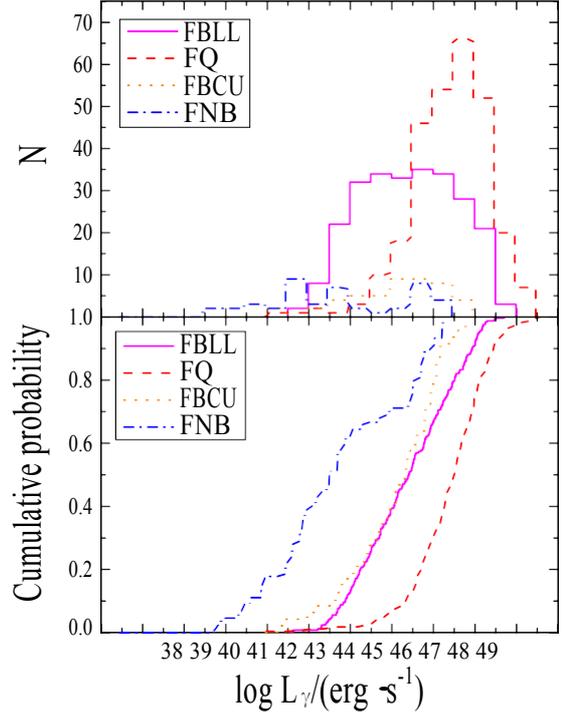

**Fig. 9.** Distribution of gamma-ray spectral index, $\alpha_\gamma$ (upper panel) and cumulative probability (lower panel) for the whole Fermi sources. In this plot, the representations of all symbols are the same as in Fig.2.

**Fig. 10.** Distribution of gamma-ray luminosity, log $L_\gamma$ (upper panel) and cumulative probability (lower panel) for the whole Fermi sources. In this plot, the representations of all symbols are the same as in Fig.2.

$\alpha^{ph}_{\gamma,core}$ is close to the statistical averaged value of photon index ($\bar{\alpha}_{ph}$) respect to the extended component $\alpha^{ph}_{\gamma,ext.}$ (see Tab.5).

Wu et al (2014a) obtained that the ratio of of γ-ray luminosity to the extended radio luminosity, log ($L_\gamma/L_{ext.}$) is associated as log $R$ for 124 Fermi blazars with a positive linear regression and proposed that "does that mean γ-ray luminosity consists of two components as do the radio bands?" As the result of Fig.17, we do not obtain a clear positive correlation between γ-ray luminosity and log $R$. Thus, we consider the two components model in the γ-ray emissions as same as in the radio band, namely, $L_{Total} = L_{core} + L_{ext.} = R\, L_{ext.} + L_{ext.} = (1 + R)L_{ext.}$, and we obtain that a clear positive correlation between the ratio of γ-ray luminosity to the extended radio luminosity, log ($L_\gamma/L_{ext.}$), and log ($R + 1$), as log ($L_\gamma/L_{ext.}$) = (0.92 ± 0.05) log ($R + 1$) + (2.68±0.06) with a correlation coefficient $r = 0.56$ and a chance probability of $p = 0.00$ (see Fig.18). Thus, if we exclude the effect of radio extended luminosity from γ-ray luminosity, then we can have a clear positive tendency between log ($L_\gamma/L_{ext.}$) and core-dominance parameter.

Therefore, if we also consider the two-component model on γ-ray emissions that composed of core (beamed) component and extended (unbeamed) one, we can claim that the γ-ray emissions of blazars is mainly from the core (jet) component (also see Pei et al.2016).

From the previous studies, the core-dominance parameter $R$ can take the role of the indicator of beaming effect (see Urry & Padovani1995; Fan2003), where $f$ is the intrinsic ratio, defined by the intrinsic flux density in the jet to the extend flux density in the co-moving frame, $f = \frac{S^{in}_{core}}{S^{in}_{ext.}}$ (Fan & Zhang2003), $\theta$ the viewing angle, $\gamma$ the Lorentz factor, $\gamma = (1 - \beta^2)^{-1/2}$, $\alpha$ is the radio spectral index and $n$ de-

$$R(\theta) = f\gamma^{-n}[(1 - \beta\cos\varphi)^{-n+\alpha} + (1 + \beta\cos\varphi)^{-n+\alpha}], \quad (11)$$

pends on the shape of the emitted spectrum and the physical de- tail of the jet, $n = 2$ for continuous jet and $n = 3$ for blobs. Therefore, $R$ is a good statistical measure and indicator of the relativistic beaming effect.

The association between the spectral index $\alpha_R$ and core- dominance parameter $R$ in radio sources is a subject of continued study. Fan et al. (2011) calculated the core-dominance parameter and the radio radio spectral index for the whole sample, and gave the relationship between $\alpha_R$ and log $R$, which indicates the spec- tral index is associated with core-dominance parameter. We also suggest that the relativistic beaming effect may result in an as- sociation between spectral index and core-dominance parameter extragalactic sources in radio emission (also see Pei et al. 2016). In the two component beaming model, the relative prominence of the core with respect to the extended emission defined as the ratio of core-to-extended-flux density measured in the rest frame of the source log $R$ has become a suitable statistical measure of beaming and orientation. The log $R$ as a beaming indicator and radio source orientation is predicated on Doppler beaming ef- fects in the radio core which is the unresolved base of the rela- tivistic jets.

Following the classification by Fanaroff & Riley (1974) and corresponding to edge-darkened and edge-brightened objects, two radio morphological subclasses named FR type one radio galaxies and type two ones. FRII objects are more powerful than

**Table 3.** Averaged values for the whole sample

| Sample (1) | N (2) | (log R) (3) | (α_R) (4) | (α_γ^ph) (5) | (log L_γ) (6) |
|---|---|---|---|---|---|
| Total | 4435 | -0.29 | 0.44 | ... | ... |
| FS | 696 | 0.53 | 0.12 | 2.29 | 45.65 |
| non-FS | 3741 | -0.44 | 0.50 | ... | ... |
| FBLL | 274 | 0.66 | 0.15 | 2.04 | 45.34 |
| non-FBLL | 195 | 0.46 | 0.26 | ... | ... |
| FQ | 306 | 0.70 | -0.03 | 2.50 | 46.40 |
| non-FQ | 1099 | 0.03 | 0.28 | ... | ... |
| FG | 32 | -0.58 | 0.64 | 2.22 | 43.32 |
| non-FG | 1425 | -0.71 | 0.62 | ... | ... |
| FSy | 17 | -0.09 | 0.32 | 2.49 | 44.36 |
| non-FSy | 507 | -0.31 | 0.61 | ... | ... |
| FFR | 9 | -0.92 | 0.69 | 2.30 | 44.53 |
| non-FFR | 340 | -1.64 | 0.85 | ... | ... |
| FBCU | 56 | 0.05 | 0.37 | 2.30 | 44.97 |
| non-FU | 175 | -0.20 | 0.28 | ... | ... |
| FB | 636 | 0.63 | 0.08 | 2.28 | 45.82 |
| non-FB | 1294 | 0.09 | 0.28 | ... | ... |

**Notes.** Col. 1 gives the sample: FS and non-FS for Fermi and non-Fermi sources, FBLL and non-FBLL for Fermi and non-Fermi BL Lacs, FQ and non-Q for Fermi and non-Fermi quasars, FG and non-G for Fermi and non-Fermi normal galaxies, FSy and non-FSy for Fermi and non-Fermi Seyfert galaxies, FFR and non-FFR for Fermi and non-Fermi FR type I and II galaxies, FBCU for Fermi blazars candidate uncertainties, non-FU for non-Fermi unidentified sources, FB and non-FB for Fermi and non-Fermi blazars; Col. 2 numbers; Col.3 averaged of core-dominance parameter; Col. 4 averaged of radio spectral index; Col. 5 averaged of γ-ray photon index for Fermi sources and Col. 6 averaged of γ-ray luminosity for Fermi sources in the unit of erg s$^{-1}$.

FRI radio galaxies. For the FRIs, the jets are thought to decelerate and become subrelativistic on scales of hundreds of pc to kpc. On the contrary, the jets in FRIIs are moderately relativistic and supersonic from the core to the hot spots. Considering the FRIs and FRIIs radio galaxies, it appears that FRIs with large jet inclination angles and correspondingly small core-dominance parameter can be observed only if they have large radio flux densities. Thus, the source flux becomes weaker with the distance increases, and made log $R$ decreases. However, there are no FRIIs observed at large angles to the jet axis as reflected by a small core-dominance parameter has yet been detected with $Fermi$/LAT. In our FR radio galaxies sample, we obtain from the redshift distribution that $(z)|_{FRI} = 0.11 \pm 0.43$ and $(z)|_{FRII} = 0.66 \pm 0.58$, therefore, FRIIs by detected with small log $R$ would simply be a consequence of FRIIs being at larger redshifts than FRIs and therefore too weak to be detected (also see Abdo et al.2010c).

Concerning the γ-ray flux density from Dermer(1995), we have that $S_\gamma \propto \delta^{3+\alpha}$ and $S_\gamma \propto \delta^{4+2\alpha}$ for the SSC model and EC model, respectively. These indices (3 + α and 4 + 2α are true for transient emission features, whereas in a steady jet, the indices are less by one, which is 2 + α and 3 + 2α. When we take the γ-ray luminosity into account, we should expect that: $L_\gamma \propto \delta^{4+\alpha}$ for the SSC model, and $L_\gamma \propto \delta^{5+2\alpha}$ for the EC model.

We found that the γ-ray luminosity is associated with the core-dominance parameter (also see Fan et al.2010), the γ-ray variability index is correlated with that in the radio band (Fan et al.2002) and also associated with the core-dominance parameter (Pei et al.2016), and there is a real correlation between the γ-ray luminosity and the Doppler factor (Fan et al.2013). All of these suggest a beaming effect in the γ-ray emissions.

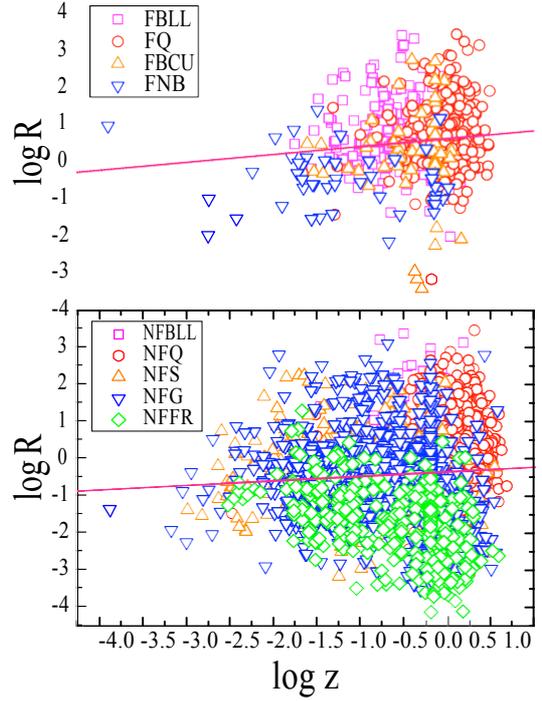

**Fig. 11.** Plot of the core-dominance parameter (log $R$) against the redshift (log $z$) for the $Fermi$ AGNs (upper panel) and non-$Fermi$ AGNs (lower panel). In the upper panel, magenta □ stands for $Fermi$ BL Lacs, red § for $Fermi$ quasars, orange for $Fermi$ BCUs, blue for $Fermi$ non-blazars. In the lower panel, magenta □ stands for non-$Fermi$ BL Lacs, red for non-$Fermi$ quasars, orange for non-$Fermi$ Seyfert galaxies, blue o for $Fermi$ normal galaxies, green § for non-$Fermi$ FRI & FRII type galaxies and grey × for non-$Fermi$ candidate uncertainty objects. The linear fitting gives that log $R = (0.21 \pm 0.08)$ log $z + (0.62 \pm 0.05)$ and log $R = (0.12 \pm 0.03)$ log $z - (0.38 \pm 0.02)$ for $Fermi$ AGNs and non-$Fermi$ AGNs, respectively.

## 4. Conclusion

From our discussions, given the the core-dominance parameter, log $R$ and radio spectral index, $\alpha_R$, we can get the $\alpha_{core}$ and $\alpha_{ext.}$. When we adopt the same performance on the γ-ray photon index, $\alpha_\gamma^{ph}$, $\alpha_{\gamma core}^{ph}$ and $\alpha_{\gamma ext.}^{ph}$ are also obtained. In this paper, we compiled 4388 AGNs with available log $R$, which included 630 $Fermi$-detected AGNs and 3758 non-$Fermi$-detected AGNs. This sample is still not sufficient enough, but we think it is adequate for the statistical analysis. Therefore, we draw the following conclusions:

1. Core-dominance parameters (log $R$) and radio spectral indices ($\alpha_R$) are quite different for different subclasses of both Fermi sources and non-Fermi sources. Particularly, Fermi sources have on average higher log $R$ than non-Fermi sources, whereas the former have on average $\alpha_R$ less than the later ones. For log $R$, we can obtain a sequence for Fermi sources as: (log $R$)|_{Fermi quasars} > (log $R$)|_{Fermi BLLacs} > (log $R$)|_{Fermi blazars candidate uncertainties} > (log $R$)|_{Fermi Seyfert galaxies} >

**Table 4.** Statistical K-S Test Results

| | Sample: A ~ B | $N_A$ | $N_B$ | $d_{max}$ | $p$ |
|---|---|---|---|---|---|
| $\log R$ | FS ~ non-FS | 696 | 3741 | 0.41 | $2.77 \times 10^{-86}$ |
| | FBLL ~ non-FBLL | 274 | 195 | 0.17 | 0.00 |
| | FQ ~ non-FQ | 306 | 1099 | 0.38 | $5.00 \times 10^{-31}$ |
| | FB ~ non-FB | 636 | 1294 | 0.30 | $1.05 \times 10^{-33}$ |
| $\alpha_R$ | FS ~ non-FS | 676 | 3177 | 0.39 | $9.20 \times 10^{-73}$ |
| | FBLL ~ non-FBLL | 265 | 179 | 0.16 | 0.01 |
| | FQ ~ non-FQ | 299 | 1066 | 0.33 | $9.91 \times 10^{-22}$ |
| $\alpha_\gamma^{ph}$ | FB ~ non-FB | 619 | 1185 | 0.23 | $5.95 \times 10^{-19}$ |
| | FBLL ~ FQ | 274 | 299 | 0.76 | $4.85 \times 10^{-73}$ |
| | FBLL ~ FBCU | 274 | 56 | 0.37 | $5.65 \times 10^{-6}$ |
| | FQ ~ FBCU | 299 | 56 | 0.43 | $6.43 \times 10^{-8}$ |
| $\log L_\gamma$ | FBLL ~ FQ | 274 | 299 | 0.43 | $4.88 \times 10^{-24}$ |
| | FBLL ~ FBCU | 274 | 56 | 0.15 | 0.26 |
| | FQ ~ FBCU | 299 | 56 | 0.52 | $1.53 \times 10^{-11}$ |

**Table 5.** Statistical results for γ-ray photon spectral indices ($\alpha_\gamma^{ph}$) against core-dominance parameters ($\log R$) for the whole *Fermi* sources

| Sample | N | $\alpha_{\gamma, core}^{ph}$ | $\alpha_{\gamma, ext.}^{ph}$ | $\chi^2$ | $R^2$ | $(\alpha_\gamma^{ph})$ |
|---|---|---|---|---|---|---|
| FS | 696 | 2.37 | 1.79 | 0.10 | $< 10^{-4}$ | 2.29 |
| FBLL | 274 | 2.05 | 1.64 | 0.05 | $< 10^{-3}$ | 2.04 |
| FQ | 306 | 2.50 | 1.99 | 0.04 | $< 10^{-3}$ | 2.50 |
| FG | 32 | 1.96 | 2.34 | 0.09 | 0.10 | 2.22 |
| FSy | 17 | 2.52 | 2.31 | 0.08 | 0.06 | 2.49 |
| FFR | 9 | 2.24 | 2.31 | 0.06 | 0.13 | 2.30 |
| FBCU | 56 | 2.34 | 1.66 | 0.09 | $< 10^{-3}$ | 2.30 |
| FB | 636 | 2.30 | 1.60 | 0.10 | $< 10^{-4}$ | 2.28 |

**Notes.** The representations for Col. 1: FS for Fermi sources, FBLL for Fermi BL Lacs, FQ for Fermi quasars, FG for Fermi galaxies, FSy for Fermi Seyfert galaxies, FFR for Fermi FR type I & type II galaxies, FBCU for Fermi blazars candidates uncertainties and FB for Fermi blazars, respectively; $\alpha_{\gamma, core}^{ph}$ and $\alpha_{\gamma, ext.}^{ph}$ represents the fitting results for $\alpha_\gamma^{ph}$ in core and extended components; $(\alpha_\gamma^{ph})$ for the statistical averaged values of γ-ray photon indices for our different sample.

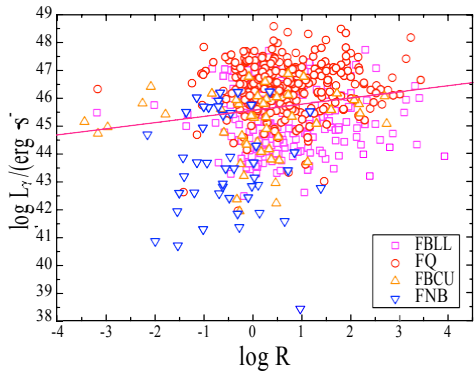

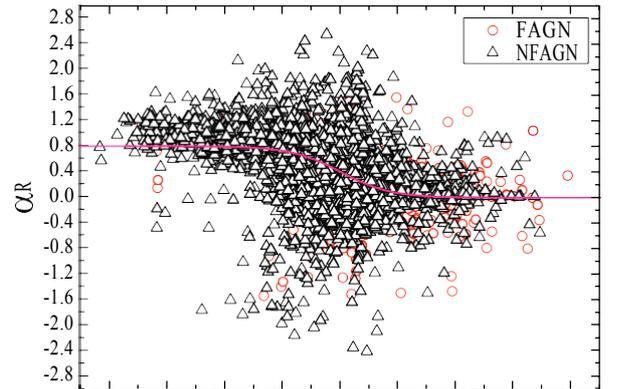

**Fig. 12.** Plot of the gamma-ray luminosity ($\log L_\gamma$) against the core dominance parameter ($\log R$) for *Fermi* AGNs. The fitting results gives that $\log L_\gamma = (0.22 \pm 0.05) \log R + (45.57 \pm 0.06)$.

dices ($\alpha_R$) and core-dominance parameters ($\log R$) are adopted

$(\log R)|_{Fermi\ galaxies} > \log R_{Fermi\ FR\ type\ I/II\ galaxies}$, and also $(\log R)|_{non\ Fermi\ BLLacs} > (\log R)|_{non\ Fermi\ quasars} > \log R)_{non -Fermi\ Seyfert\ galaxies} > (\log R)|_{non- Fermi\ galaxies} > \log R)|_{non\ Fermi\ FR\ type\ I/II\ galaxies}$ is also obtained for non-Fermi sources.

2. A theoretical correlation fittings between radio spectral in-

| | | | | | | | | |
|---|---|---|---|---|---|---|---|---|
| -4 | -3 | -2 | -1 | 0 | 1 | 2 | 3 | 4 |

log R

**Fig. 13.** Plot of the radio spectral index ($\alpha_R$) against the core-dominance parameter (log $R$) for the whole sample. In this plot, red § stands for *Fermi* AGNs, black a for non-*Fermi* AGNs. The fitting curve corre- sponds to $\alpha_{core} = -0.021$ and $\alpha_{ext.} = 0.785$.

for both Fermi sources and non-Fermi sources, and also obtained for all subclasses, which means the spectral index is dependent

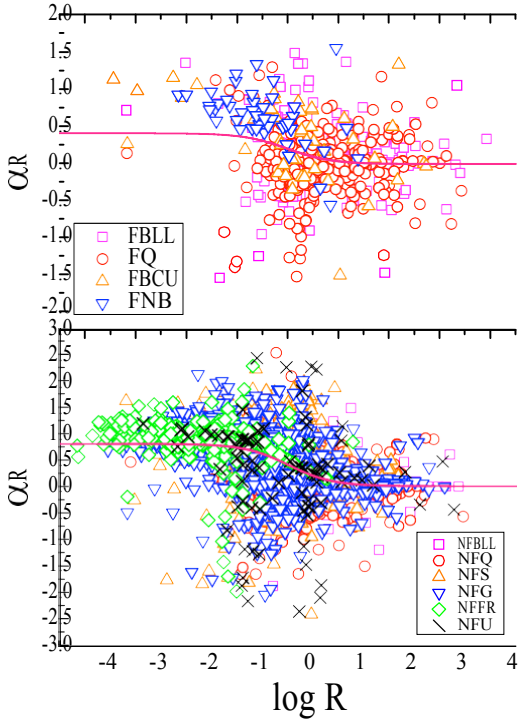
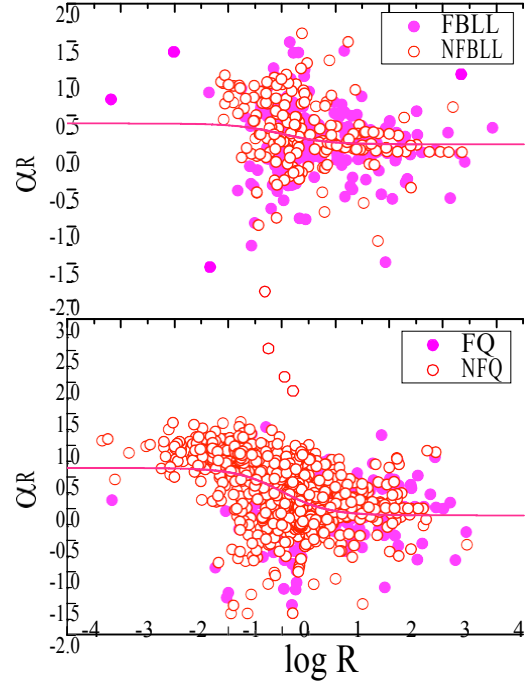

**Fig. 14.** Plot of the radio spectral index ($\alpha_R$) against the core-dominance parameter (log $R$) for the *Fermi* AGNs (upper panel) and non-*Fermi* AGNs (lower panel). In this plot, all symbols' meanings are same with Figure.11. The fitting curve corresponds to $\alpha_{core} = -0.012$ and $\alpha_{ext.} = 0.402$ (upper panel), $\alpha_{core} = 0.006$ and $\alpha_{ext.} = 0.805$ (lower panel) for *Fermi* AGNs and non-*Fermi* AGNs, respectively.

**Fig. 15.** Plot of the radio spectral index ($\alpha_R$) against the core-dominance parameter (log $R$) for the all BL Lacs (FBLL + NFBLL) and quasars (FQ + NFQ). In the upper panel, magenta solid §s stands for *Fermi* BL Lacs and red empty§ for non-*Fermi* BL Lacs; in the lower panel, magenta solid §stands for *Fermi* quasars and red empty§ for non-*Fermi* quasars. The fitting curve corresponds to $\alpha_{core} = 0.104$ and $\alpha_{ext.} = 0.385$ for BL Lacs (upper panel) and $\alpha_{core} = 0.107$ and $\alpha_{ext.} = 0.643$ for quasars (lower panel).

on the core-dominance parameter, probably from the relativistic beaming effect.

3. There is a trend that the larger log $R$ also comes the larger γ-ray luminosity ($L_\gamma$).

4. The correlation fittings between γ-ray photon spectral indices ($\alpha_\gamma^{ph}$) and core-dominance parameters (log $R$) are also taken into account for Fermi sources. if we consider that γ-ray emissions are also composed of two components, and the emissions are mainly from the core (jet) component.

5. Fermi blazars are beamed.

*Acknowledgements.* This work is partially supported by the National Natural Science Foundation of China (11733001, U1531245, NSFC 10633010, NSFC 11173009), Natural Science Foundation of Guangdong Province (2017A030313011), and supports for Astrophysics Key Subjects of Guangdong Province and Guangzhou City. Author Z. -Y. Pei would like to dedicated this paper to my wife, Sim Liu, and our new born baby, Kingsley. Daddy loves you all.

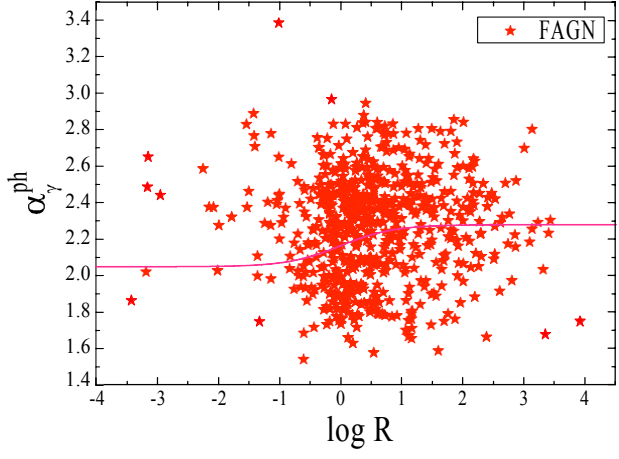

**Fig. 16.** Plot of the gamma-ray photon spectral index ($\alpha_\gamma^{ph}$) against the core-dominance parameter (log $R$) for the whole *Fermi* AGNs. The fitting curve corresponds to $\alpha_{\gamma,\,core}^{ph}$ = 2.279 and $\alpha_{\gamma,\,ext.}^{ph}$ = 2.049.

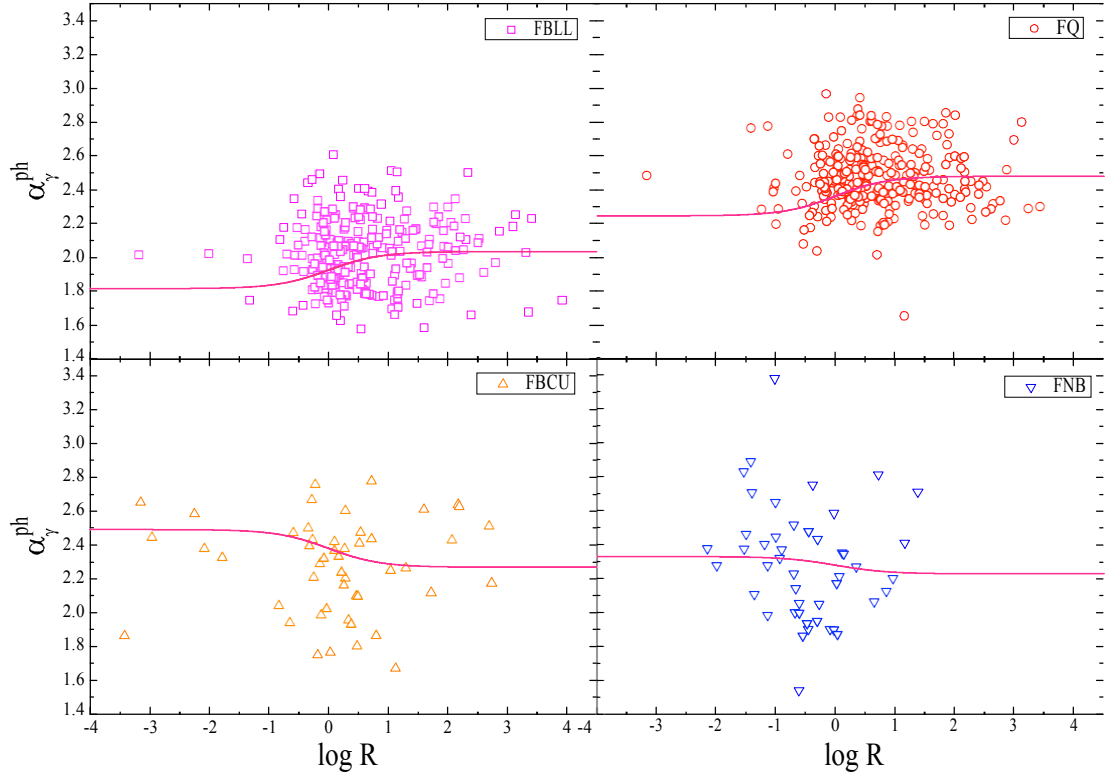

**Fig. 17.** Plot of the gamma-ray photon spectral index ($\alpha^{ph}_\gamma$) against the core-dominance parameter (log $R$) for the *Fermi* BL Lacs (upper left panel), *Fermi* quasars (upper right panel), *Fermi* blazars candidate uncertainty objects (lower left panel) and *Fermi* non-blazars (lower right panel). The fitting curve corresponds to $\alpha^{ph}_{\gamma,core} = 2.039$ and $\alpha^{ph}_{\gamma,ext.} = 1.821$ for *Fermi* BL Lacs, $\alpha^{ph}_{\gamma,core} = 2.482$ and $\alpha^{ph}_{\gamma,ext.} = 2.249$ for *Fermi* quasars, $\alpha^{ph}_{\gamma,core} = 2.266$ and $\alpha^{ph}_{\gamma,ext.} = 2.488$ for *Fermi* blazars candidate uncertainty objects, $\alpha^{ph}_{\gamma,core} = 2.227$ and $\alpha^{ph}_{\gamma,ext.} = 2.329$ for *Fermi* non-blazars, respectively.

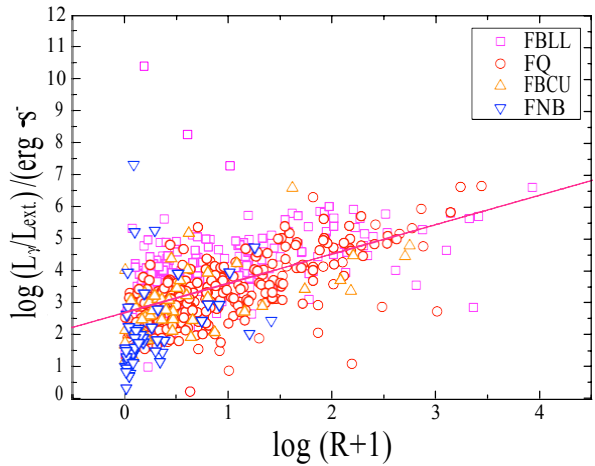

**Fig. 18.** Plot of the ratio of γ-ray luminosity to the extended radio luminosity, log ($L_\gamma/L_{ext.}$), against the core-dominance parameter (log $R$) for the *Fermi* AGNs. The fitting results gives that log ($L_\gamma/L_{ext.}$) = $(0.92 \pm 0.05) \log(R + 1) + (2.68 \pm 0.06)$.